\documentclass{article}
\usepackage{spconfa4,amsmath,graphicx}
\usepackage{bm}
\usepackage{amsmath}
\usepackage{amssymb}
\usepackage{adjustbox}
\usepackage{multirow}
\usepackage{subfigure}
\usepackage{subcaption}
\usepackage[colorlinks,
            linkcolor=black,  
            anchorcolor=black,
            urlcolor=black,
            citecolor=black,  
            ]{hyperref}
\title{TOWARDS BALANCED SPECTRAL RECONSTRUCTION: SPECTRALLY ADAPTIVE LOSS FOR STREAMING SPEECH ENHANCEMENT
}
\name{Haixin Zhao and Nilesh Madhu}
\address{IDLab, Department of Electronics and Information Systems, Ghent University - imec, Ghent, Belgium\\
\{haixin.zhao, nilesh.madhu\}@ugent.be
}
\begin{document}
%
\maketitle
\begin{abstract}
This paper proposes two spectrally weighted STFT loss functions for lightweight streaming speech enhancement, addressing the magnitude over-attenuation in mid-to-high frequency regions caused by the magnitude-phase compensation effect. The proposed sigmoid-weighted loss applies a smooth frequency-dependent modulation to the phase-aware contribution, while the signal-dependent spectrally adaptive loss further conditions the modulation on the ground-truth log-magnitude spectrogram. To evaluate the proposed objectives, we additionally design HyST-Net, a lightweight and competitive backbone with hybrid MHA-GRU spectral-temporal modelling for low-latency streaming scenarios. Experimental results exhibit consistent improvements in high-frequency spectral reconstruction for both losses. The spectrally adaptive loss further enhances the mid-frequency region, resulting in a more balanced spectral reconstruction across the full frequency range.
\end{abstract}
%
\section{Introduction}
\label{sec:intro}

Over the past decade, numerous lightweight deep neural network architectures have been proposed for real-time and resource-constrained speech enhancement (SE) \cite{9413580, 9914782, 11226622}. 
Among these, convolutional recurrent neural networks (CRNNs) 
\cite{9413580, 9914782, wu23b_interspeech, girirajan2023real} model 
local patterns through convolutional encoders and long-term temporal 
dependencies through recurrent bottlenecks, but suffer from 
disproportionate recurrent overhead caused by the flattening of 
frequency and channel dimensions \cite{11226622}.
Dual-path RNNs \cite{9054266} were introduced to model temporal and spectral dependencies in an interleaving manner \cite{10448310, yang2024fspen, le21b_interspeech, 11071541}, reducing sequential modelling complexity via parameter sharing. Building on this, FTF-Net incorporates multi-head attention (MHA~\cite{NIPS2017_3f5ee243}) blocks alongside RNN units for joint, interleaved modelling \cite{11226622, zhao2026dynamicallyslimmablespeechenhancement}. 
However, RNN-based spectral modelling lacks parallelism, while causal MHA for temporal modelling incurs a per-frame computational cost that grows linearly with context length even with key-value caching \cite{shazeer2019fast}. Both impose non-trivial overhead for real-time streaming inference on edge devices.

Moreover, the stringent latency and computational constraints of lightweight streaming SE scenarios bound model capacity. As compact models offer limited representational capacity to compensate for biases introduced by a suboptimal loss function, the design of the training objective thus becomes a critical consideration.
The compressed phase-aware short-time Fourier transform (STFT) loss has been widely adopted in SE methods \cite{10.1145/3197517.3201357, 8521347}. 
However, this formulation is susceptible to a magnitude-phase compensation effect \cite{9552504}, characterised by the network’s tendency to suppress magnitudes in regions with unreliable phase predictions. 
This results in inaccurate magnitude estimates that manifest as energy attenuation, particularly pronounced in mid-to-high frequency components, thereby degrading perceptual brightness.

The typical approach is to supplement the phase-aware loss with an additional magnitude-domain term \cite{9413580, 9914782, 11226622, 9522648}. However, as spectral energy attenuation predominantly occurs in mid-to-high frequency regions, a uniform weighting across the spectrogram inevitably trades off noise suppression against perceptual fidelity, and thus reflects an inherent compromise rather than a principled solution.

\textbf{Contributions:} 
To address the trade-off between over-attenuation at high frequencies and noise suppression at low frequencies caused by the compensation effect, we propose two spectrally weighted STFT objectives that modulate the phase-aware loss contribution as a function of frequency and signal characteristics.
The sigmoid-weighted loss $\mathcal{L}_\mathrm{Sig}$ applies a fixed frequency-dependent sigmoid weighting to suppress the phase-aware contribution at higher frequencies, while the spectrally adaptive loss $\mathcal{L}_\mathrm{Adp}$ derives a signal-dependent frequency-wise weight from the clean log-magnitude spectrogram. 
To evaluate the proposed objectives, we further design HyST-Net, a lightweight and competitive network with hybrid spectral-temporal modelling that serves as a low-latency baseline network for streaming SE. 
Experimental results show consistent improvements in high-frequency spectral reconstruction for both losses, with $\mathcal{L}_\mathrm{Adp}$ further enhancing mid-frequency reconstruction while maintaining overall enhancement quality.

\vspace{-0.1cm}
\section{Methods}
\vspace{-0.1cm}
\subsection{Spectrally Weighted STFT Loss Functions}
  \vspace{-0.1cm}
\begin{figure}[t]
    \centering
    \setlength{\tabcolsep}{2pt}
    \begin{tabular}{ccc}
        \includegraphics[width=0.32\linewidth]{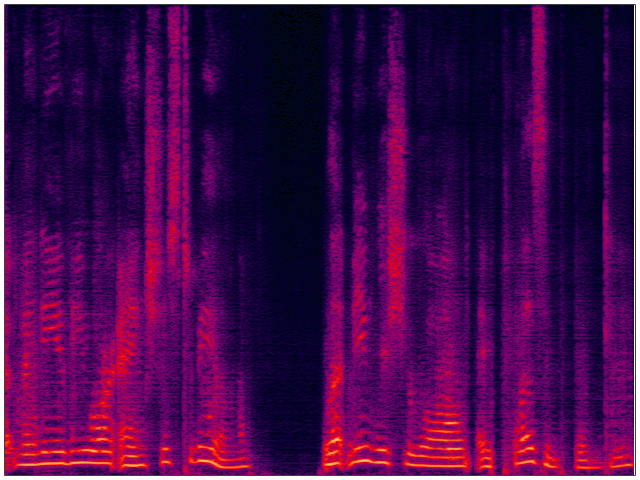} &
        \includegraphics[width=0.32\linewidth]{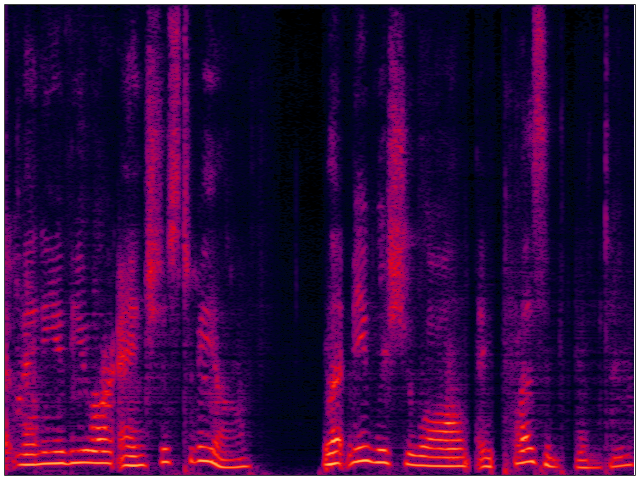} &
        \includegraphics[width=0.32\linewidth]{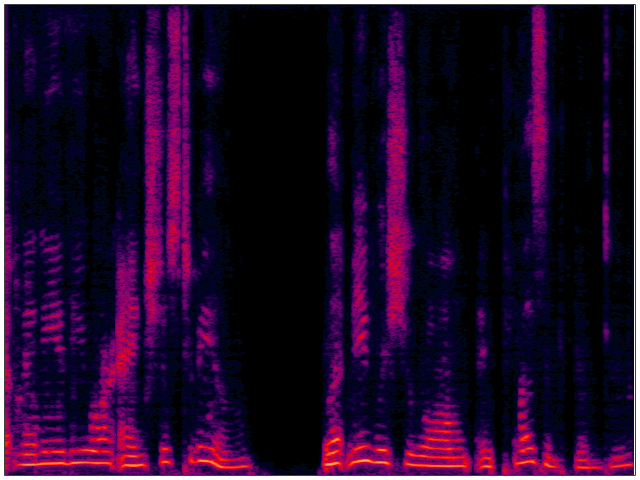} \\[-2pt]
        \small (a) $\lambda=0$ & \small (b) $\lambda=0.3$ & \small (c) $\lambda=1$ \\
    \end{tabular}
    \vspace{-4pt}
        \vspace{-0.2cm}
    \caption{Illustrative spectrograms of enhanced speech under varying phase-aware loss weights $\lambda \in \{0,0.3,1\}$. A larger $\lambda$ improves noise suppression but causes over-attenuation in mid-to-high frequencies due to the compensation effect.}
    \label{fig:0}
        \vspace{-0.3cm}
\end{figure}

The phase-aware compressed loss, originally proposed in \cite{10.1145/3197517.3201357, 8521347}, has been widely adopted in mainstream lightweight SE methods \cite{9413580, 9914782, 11226622}. As demonstrated in \cite{9522648}, combining a phase-aware component with a magnitude-only loss yields improvements in instrumental metrics. The combined loss is formulated as:
\begin{equation}
\begin{aligned}
\textstyle
    \mathcal{L}_{\mathrm{Mix}} &= (1-\lambda) \mathcal{L}_\mathrm{Mag} + \lambda \mathcal{L}_\mathrm{Pha} \\
&=    (1-\lambda) \lvert\bm{\widehat{S}}^c - \bm{S}^c\rvert^{2} + \lambda \lvert\bm{\widehat{S}}^c exp^{j\bm{\phi_{\widehat{S}}}} - \bm{S}^c exp^{j\bm{\phi_{S}}}\rvert^{2}
\end{aligned}
    \label{equation:eq2}
\end{equation}
where $\mathcal{L}_\mathrm{Mag}$ and $\mathcal{L}_\mathrm{Pha}$ correspond to the compressed magnitude loss and the phase-aware loss, respectively. 
$S$ and $\widehat{S}$ denotes the ground-truth and estimated spectrogram, and $c=0.3$ is the power-compression factor.
$\lambda$ controls the trade-off between noise suppression and spectral reconstruction. 
As illustrated in Fig.~\ref{fig:0} (a) and (c), an insufficient $\lambda$ leads to inadequate noise suppression, while an excessive $\lambda$ causes pronounced over-attenuation, particularly in mid-to-high frequency regions. This arises from the magnitude-phase compensation effect \cite{9552504}: when phase estimation is unreliable, the phase-aware loss drives the estimated magnitude toward zero to minimise its value, resulting in over-attenuation. To balance this trade-off, $\lambda = 0.3$ has been reported to yield the best instrumental metric performance \cite{9522648}.

The compensation effect empirically exhibits spectrally non-uniform over-attenuation across frequency, predominantly concentrated in mid-to-high frequency regions \cite{11226622}.
The scalar weighting $\lambda$ therefore does not account for this spectral variation.
To address this limitation, we propose a sigmoid-weighted loss $\mathcal{L}_\mathrm{Sig}$ that applies a frequency-dependent weight to the phase-aware component:
\begin{equation}
\textstyle
    \mathcal{L}_\mathrm{Sig} = 0.7 \cdot \mathcal{L}_\mathrm{Mag} + \lambda_{\mathrm{sig}}  \cdot \sigma(\beta  \cdot (f_{\mathrm{n}}-r))  \cdot \mathcal{L}_\mathrm{Pha}
    \label{equation:eq3}
\end{equation}
where $\sigma$ denotes the sigmoid function. $f_n \in [0, 1]$ is the normalised frequency variable with $f_n = 1$ corresponds to the Nyquist frequency bin. The cut-off ratio $r$, weighting coefficient $\lambda_{sig}$, and steepness $\beta$ are empirically set to 0.4, 0.5, and -20, respectively. By introducing a smooth frequency-dependent transition, $\mathcal{L}_\mathrm{Sig}$ suppresses the phase-aware contribution in mid-to-high frequency regions where over-attenuation is more severe, while avoiding spectral banding artefacts.

\begin{figure}[tbp]
    \centering
    \setlength{\tabcolsep}{2pt}
    \begin{tabular}{ccc}
        \includegraphics[width=0.46\linewidth]{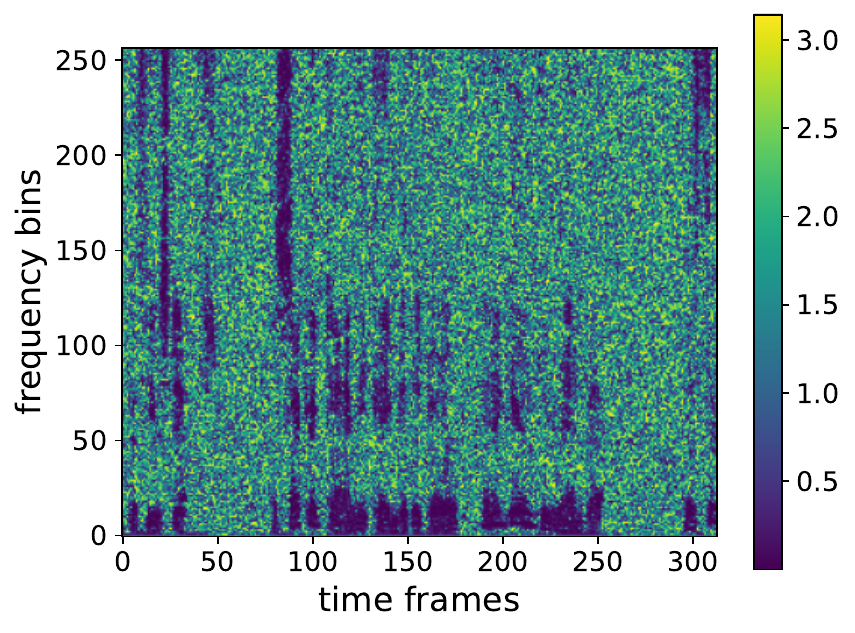} &
        \includegraphics[width=0.48\linewidth]{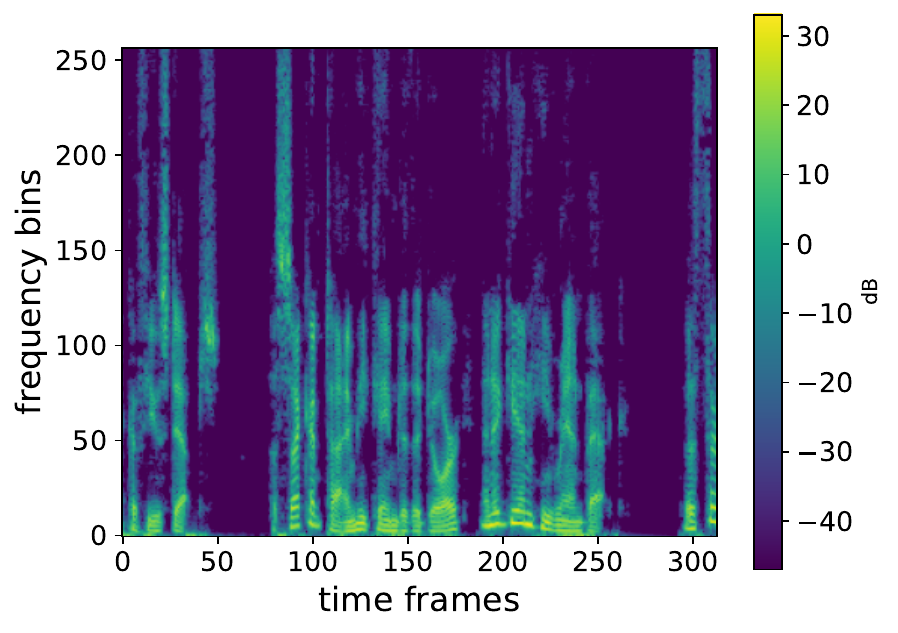}  \\[-2pt]
        \small (a) phase estimation error & \small (b) ground-truth magnitude \\
    \end{tabular}
    \vspace{-4pt}
        \vspace{-0.1cm}
    \caption{Phase estimation error and corresponding ground truth log-magnitude spectrogram of an example utterance. A notable correlation is observed between the two.}
    \label{fig:02}
        \vspace{-0.5cm}
\end{figure}

Nonetheless, $\mathcal{L}_\mathrm{Sig}$ applies a fixed frequency-dependent weighting uniformly across all utterances, regardless of signal characteristics.
Empirically, phase estimation accuracy exhibits a notable correlation with spectral magnitude. Regions with higher magnitude tend to yield more accurate phase predictions, while low-magnitude regions are more susceptible to phase estimation errors, as illustrated in Fig.~\ref{fig:02}.
To further address these limitations, we propose a signal-dependent spectrally adaptive loss $\mathcal{L}_\mathrm{Adp}$, where the ground-truth log-magnitude spectrogram is used to derive a spectrally adaptive weight that modulates the phase-aware loss contribution:
\begin{equation}
\textstyle
    \mathcal{L}_\mathrm{Adp} = 0.7  \cdot \mathcal{L}_\mathrm{Mag} + \lambda_{\mathrm{adp}} \cdot \mathcal{F}_{s}(\sigma(\mathcal{N}(\mathbb{E}_t[\log \lvert\bm{S} \rvert])))
     \cdot\mathcal{L}_\mathrm{Pha}
    \label{equation:eq4}
\end{equation}
where $\mathcal{F}_{s}(\cdot)$ is a 1D spectral smoothing operator, and $\mathcal{N}(\cdot)$ is min-max normalisation. $\mathbb{E}_t[\cdot]$ represents averaging along the time dimension. 
The coefficient $\lambda_{\mathrm{adp}}$, sigmoid steepness and cut-off are empirically set to 0.6, 15 and 0.5, respectively. The steepness is smaller relative to $\mathcal{L}_\mathrm{Sig}$, as the sigmoid here operates on log-magnitude rather than frequency.

\begin{figure*}[t]
  \centering
  \includegraphics[width=0.8\linewidth]{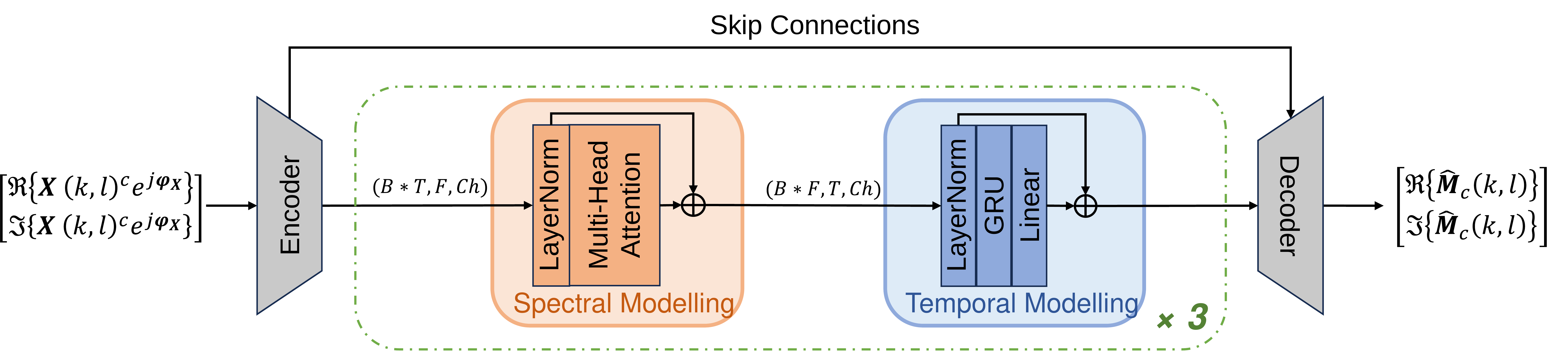}
    \vspace{-0.3cm}
  \caption{Architectures of the proposed hybrid spectral-temporal modelling network (HyST-Net). In the bottleneck, $B$ denotes the batch size, and $T$, $F$, and $Ch=64$ are the tensor sizes along time, frequency and channel, respectively.
  }
  \vspace{-0.4cm}
  \label{fig:fig1}
\end{figure*}

  \vspace{-0.1cm}
\subsection{Streaming HyST-Net}
  \vspace{-0.1cm}
To evaluate the proposed objectives under practical streaming scenarios, we further design HyST-Net as a lightweight backbone. The architecture is shown in Fig.~\ref{fig:fig1}. 
The real and imaginary parts of the compressed noisy complex spectrogram $\bm{X}(k,l)$ are concatenated channel-wise as input. HyST-Net estimates a complex-valued ideal ratio mask 
$\bm{\widehat{M}}_{c}(k,l)$ in the compressed domain, defined as:
\begin{equation}
    \bm{\widehat{M}}_{\text{c}}(k,l) = \frac{|\bm{S}(k,l)|^{c} exp^{j\bm{\phi_{S}}(k,l)}} {|\bm{X}(k,l)|^{c} exp^{j\bm{\phi_{X}}(k,l)} + \gamma}
\end{equation}
where $k,l$ index the frequency bin and time frame, respectively. $|\bm{S}(k,l)|^{c} exp^{j\bm{\phi_{S}}(k,l)}$ denotes the compressed clean reference spectrogram, and $\gamma$ is a small regularisation constant for numerical stability. The enhanced complex spectrogram $\bm{\widehat{S}}(k,l)$ is subsequently recovered by applying the estimated mask followed by magnitude decompression.

HyST-Net adopts a U-Net architecture with a three-layer causal convolutional encoder–decoder configured identically to FTF-Net \cite{11226622}, with one-time-step buffer caches in convolutional layers to support streaming inference.
In the bottleneck, HyST-Net employs an interleaved spectral-temporal modelling strategy. Unlike prior works that apply the same sequential module, either RNN or transformer, across both dimensions \cite{10448310, yang2024fspen, 11226622}, HyST-Net tailors the module to each dimension by applying MHA for spectral modelling and GRUs for temporal modelling. This hybrid design exploits the complementary strengths of the two architectures. For spectral modelling, all frequency bins at a given time frame are simultaneously available, allowing MHA to leverage its inherent parallelism and avoid the processing latency incurred by recurrent spectral modelling. For temporal modelling, GRUs are preferred for their compact recurrent state, which captures long-range context efficiently without explicit attention over all past context time steps. Although key-value caching \cite{shazeer2019fast} reduces causal MHA complexity to linear in context length, its memory and computational overhead remain non-trivial relative to GRUs in lightweight streaming scenarios. Three interleaved spectral-temporal blocks are stacked in the bottleneck to balance enhancement performance and computational cost for real-time deployment.

  \vspace{-0.2cm}
\section{Experiments}
 \vspace{-0.15cm}

\subsection{Experiment Setup}
\label{sec:31}
  \vspace{-0.15cm}
Experiments are conducted on the DNS Challenge dataset \cite{reddy21_interspeech}. The training set comprises 140 hours of synthesised wideband speech generated from the clean speech and noise corpora provided by DNS, with SNRs ranging from $-5$ dB to $20$ dB. Evaluation is performed on the public synthetic test set from the DNS Challenge \cite{reddy20_interspeech}. The STFT is computed with a window length of 512 samples and 50\% overlap. The square-root von Hann window is used for analysis and synthesis. 
HyST-Net is configured following \cite{11226622}, and is optimised using AdamW with exponential decay rates of $(0.9, 0.99)$. All proposed loss functions are implemented in a multi-resolution form with STFT sizes $m \in \{320, 512, 768\}$ and 50\% overlap \cite{11226622}, denoted as $\mathcal{L}_\mathrm{MR\_Mix}$, $\mathcal{L}_\mathrm{MR\_Sig}$, and $\mathcal{L}_\mathrm{MR\_Adp}$.

  \vspace{-0.15cm}
\subsection{Experimental Results}
  \vspace{-0.15cm}
\begin{table*}[th]
\caption{Evaluation results of the proposed HyST-Net against lightweight baselines under the causal streaming condition}
\label{tab:table1}
\centering
\setlength{\tabcolsep}{2mm}{
\begin{adjustbox}{width=1.0\linewidth}
\begin{tabular}{ |l|ccc|cccc|} \hline
{\textbf{Model}} &
\textbf{MACs [M/s]} & \textbf{Param [M]}   & \textbf{RTF} & \textbf{DNSMOS $\uparrow$ }  & \textbf{PESQ $\uparrow$ }  & \textbf{ESTOI $\uparrow$ }   & \textbf{SI-SDR $\uparrow$ }
  \\ \hline
{Noisy speech}                 &  - & -   &  -           &  2.48 ($\pm0.49$)         & 1.58 ($\pm0.46$)   & 0.810 ($\pm0.121$) & 9.07 ($\pm5.47$)   \\ 
 \hline
CRUSE4-64-1$\times$GRU2                 &  301.2 & 2.85  &  0.26   &  3.22 ($\pm0.20$)             & 2.84 ($\pm0.66$)   & 0.912 ($\pm0.067$) & 17.19 ($\pm4.42$)   \\ 
FTF-Net   & 318.2 &  0.14   &  1.05 &  \textbf{3.23} ($\pm0.22$)     & \textbf{2.91} ($\pm0.66$)   & \textbf{0.917} ($\pm0.065$) & \textbf{17.48} ($\pm4.42$)  \\ 
 \hline

{HyST-Net}       & \textbf{266.4}  & \textbf{0.11}   &  \textbf{0.22}     &  \textbf{3.23} ($\pm0.21$)       & 2.86 ($\pm0.66$)   & 0.914 ($\pm0.068$) & 17.41 ($\pm4.46$)   \\ 

  \hline
	\end{tabular}
        \end{adjustbox}}
\end{table*}

\begin{table*}[th]
\caption{Evaluation results on proposed loss functions based on HyST-Net}
\label{tab:table2}
\centering
\setlength{\tabcolsep}{0.1mm}{
\begin{adjustbox}{width=1.0\linewidth}
\begin{tabular}{ |l|c|c|c|c|c|c|c|c|c|c|c|c|} \hline
\multirow{2}{*}{\textbf{Loss}} &  \multicolumn{4}{c|}{\textbf{Overall Metrics}}  &  \multicolumn{4}{c|}{\textbf{HF ($\bm{4-8 \ k Hz}$) Metrics} }& \multicolumn{4}{c|}{\textbf{MF ($\bm{2-4 \ k Hz}$) Metrics}}
\\
\cline{2-13}
&
\textbf{DNSMOS $\uparrow$ } & \textbf{PESQ $\uparrow$ }  & \textbf{ESTOI $\uparrow$ }   & \textbf{SI-SDR $\uparrow$ } & \textbf{C-RMSE $\downarrow$} & \textbf{M-RMSE $\downarrow$} & \textbf{LSD $\downarrow$}& \textbf{SI-SDR $\uparrow$}  & \textbf{C-RMSE $\downarrow$} & \textbf{M-RMSE $\downarrow$} & \textbf{LSD $\downarrow$}& \textbf{SI-SDR $\uparrow$}  
  \\ \hline
{Noisy}              & 2.48   &  1.58 & 0.810    &  9.07  &  0.0653         & 0.0604   & 12.16  & 10.96 &  0.1416 & 0.1288  & 12.1  &  8.31     \\ 
 \hline
 
$\mathcal{L}_\mathrm{MR\_Mix} $        & \textbf{3.23}   &  \textbf{2.86} & \textbf{0.914}    &  \textbf{17.41}  &  0.0273         & 0.0237   & 8.67  & 13.92 &  0.0539 & 0.0445  & 7.78  &  13.18    \\ 

 $\mathcal{L}_\mathrm{MR\_Sig}$   & 3.21   &  2.81 & 0.912    &  17.36  &  0.0247         & 0.0201   & \textbf{7.48}  & \textbf{14.36} &  0.0534 & 0.0452  & 7.88  &  13.24    \\ 
 
 $\mathcal{L}_\mathrm{MR\_Adp}$   & 3.21   &  2.85 & \textbf{0.914}    &  17.33  &  \textbf{0.0246}  & \textbf{0.0200}   & 7.49  & 14.35 &  \textbf{0.0522} & \textbf{0.0427}  & \textbf{7.46}  &  \textbf{13.48}    \\

  \hline
	\end{tabular}
    \end{adjustbox}
    }
    \vspace{-0.2cm}
\end{table*} 

\subsubsection{Validation of the baseline network}
    \vspace{-0.15cm}
\label{sec:321}

Table~\ref{tab:table1} compares HyST-Net against representative lightweight baselines from both the CRNN-based paradigm (CRUSE \cite{9413580}) and the interleaved-modelling paradigm (FTF-Net 
\cite{11226622}). 
All models are implemented causally, supporting frame-by-frame streaming with an algorithmic latency of 32 ms, and follow the same complex-valued compressed-domain mask formulation. They are trained with the multi-resolution baseline loss $\mathcal{L}_\mathrm{MR\_Mix}$ \cite{11226622}. For CRUSE, a variant with a bottleneck channel size of 64, one GRU, and a group factor of 2 is included as a representative of comparable computational scale.

Performance is evaluated using DNSMOS P.835 \cite{reddy2022dnsmos}, PESQ \cite{itut07}, ESTOI \cite{taal2010short}, and SI-SDR \cite{le2019sdr}, with computational overhead characterised by MACs, parameter count, and real-time factor (RTF). RTF is measured under strict frame-by-frame streaming on a single thread of Intel Xeon Silver 4310 CPU without block-wise buffering, ensuring fair comparison under minimal-latency streaming conditions. No ONNX-based optimisation is applied to the reported values.

As shown in Table~\ref{tab:table1}, HyST-Net achieves enhancement performance comparable to FTF-Net across all metrics while reducing MACs by 16.3\% and parameters by 21\%. More importantly for streaming deployment, HyST-Net attains an RTF of 0.22, approximately 4.77\,$\times$ faster than FTF-Net (RTF 1.05) on CPU. The high RTF of FTF-Net is largely attributable to the recurrent bottleneck introduced by RNN-based spectral modelling, which our hybrid design avoids through parallel MHA-based spectral processing. Compared to CRUSE, HyST-Net achieves comparable or better performance at a similar RTF with 96\% fewer parameters. These results validate HyST-Net as a competitive backbone for evaluating the proposed loss functions under practical streaming conditions.

  \vspace{-0.15cm}
\subsubsection{Evaluation of proposed loss functions}
  \vspace{-0.15cm}

\begin{figure}[t]
    \centering
    \setlength{\tabcolsep}{2pt}  
    \renewcommand{\arraystretch}{0.8}  
    \begin{tabular}{cc}
        \includegraphics[width=0.48\linewidth]{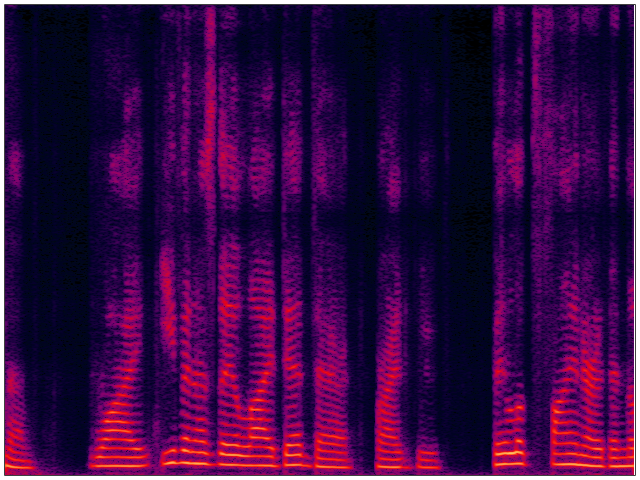} &
        \includegraphics[width=0.48\linewidth]{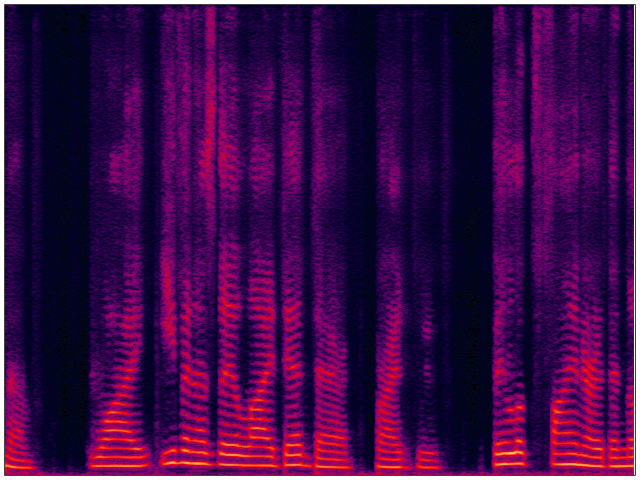} \\[-2pt]
        \small (a) $\mathcal{L}_\mathrm{MR\_Mix}$ & \small (b) $\mathcal{L}_\mathrm{MR\_Sig}$ \\[4pt]
        \includegraphics[width=0.48\linewidth]{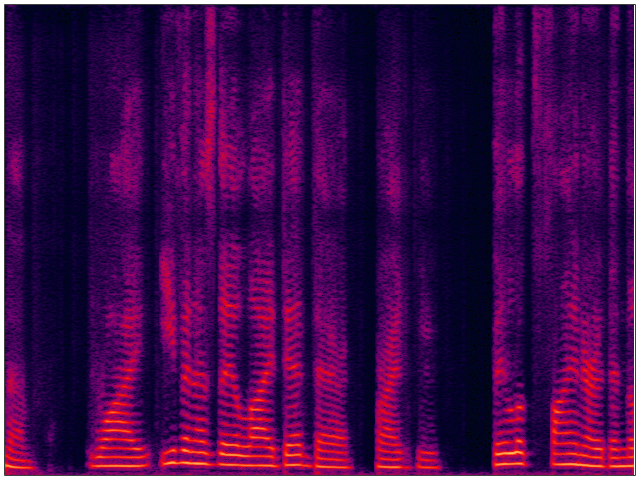} &
        \includegraphics[width=0.48\linewidth]{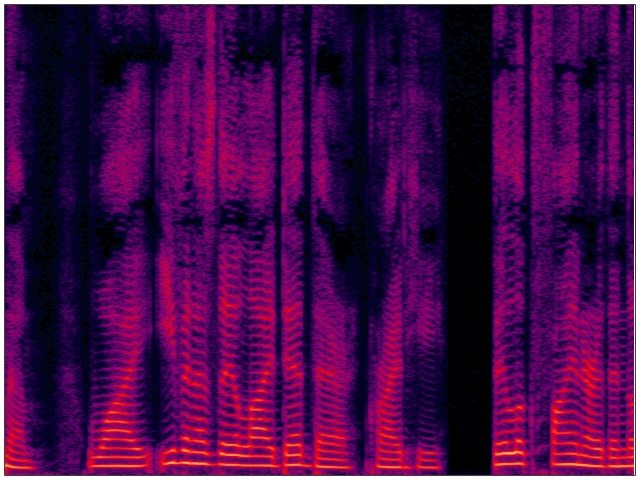} \\[-2pt]
        \small (c) $\mathcal{L}_\mathrm{MR\_Adp}$ & \small (d) clean \\
    \end{tabular}
    \vspace{-4pt}
        \vspace{-0.05cm}
    \caption{Enhanced spectrograms of an example utterance under different loss functions. $\mathcal{L}_\mathrm{MR\_Adp}$ reconstructs more energy in mid-to-high frequency regions while maintaining effective noise suppression at low frequencies.}
    \label{fig:4}
        \vspace{-0.25cm}
\end{figure}

Table~\ref{tab:table2} compares the proposed $\mathcal{L}_\mathrm{MR\_Sig}$ and $\mathcal{L}_\mathrm{MR\_Adp}$ against the baseline $\mathcal{L}_\mathrm{MR\_Mix}$, all trained on HyST-Net under the same multi-resolution configuration described in Sec.~\ref{sec:31}. As overall metrics such as PESQ, ESTOI, and DNSMOS are dominated by low-frequency components and are relatively less sensitive to distortions in mid-to-high frequencies, evaluation is further conducted in the MF (2–4 kHz) and HF (4–8 kHz) regions using complex root-mean-squared error (C-RMSE), magnitude root-mean-squared error (M-RMSE), log-spectral distance (LSD) \cite{4317555}, and SI-SDR. The first three are computed directly from spectral coefficients within each band, while SI-SDR is obtained after band-pass filtering in the time domain.
As shown in Table~\ref{tab:table2}, both $\mathcal{L}_\mathrm{MR\_Sig}$ and $\mathcal{L}_\mathrm{MR\_Adp}$ achieve overall metrics on par with $\mathcal{L}_\mathrm{MR\_Mix}$.
This is expected, as the proposed losses retain a comparably high phase-aware weighting at low frequencies, and the instrumental metrics are generally dominated by low-frequency components.

In the HF region, both proposed losses yield consistent improvements over the baseline, reducing C-RMSE and M-RMSE by approximately 9.5\% and 15.2\%, respectively, with gains of 1.18 dB in LSD and 0.43 dB in SI-SDR. These results indicate that the proposed $\mathcal{L}_\mathrm{MR\_Sig}$ and $\mathcal{L}_\mathrm{MR\_Adp}$ achieve improved spectral reconstruction in the high frequency regions by modulating the phase-aware contribution spectrally.

In the mid-frequencies, $\mathcal{L}_\mathrm{MR\_Adp}$ consistently outperforms both $\mathcal{L}_\mathrm{MR\_Sig}$ and $\mathcal{L}_\mathrm{MR\_Mix}$ across all metrics. This advantage is likely linked to the signal-dependent weighting of $\mathcal{L}_\mathrm{MR\_Adp}$, which adapts the phase-aware contribution to frequency-wise spectral energy. In contrast, the fixed sigmoid modulation in $\mathcal{L}_\mathrm{MR\_Sig}$ is signal-agnostic and may not capture such variation. The consistent improvements in C-RMSE and M-RMSE across both MF and HF regions further confirm that $\mathcal{L}_\mathrm{MR\_Adp}$ achieves enhanced spectral reconstruction across the mid-to-high frequency range. 

This is also visually confirmed in Fig.~\ref{fig:4}, which presents the enhanced spectrograms of an example utterance under each loss function alongside the clean reference. 
Compared to $\mathcal{L}_\mathrm{MR\_Mix}$ and $\mathcal{L}_\mathrm{MR\_Sig}$, $\mathcal{L}_\mathrm{MR\_Adp}$ improves the spectral reconstruction across the mid-to-high frequency regions with reduced over-attenuation, while the low-frequency harmonic structures remain clearly preserved, without introducing noticeable artefacts.
As spectrograms offer limited insight into perceptual quality, the audio samples and frequency-wise weight visualisations are available at: \href{https://aspire.ugent.be/demos/IWAENC2026HZ/}{\nolinkurl{https://aspire.ugent.be/demos/IWAENC2026HZ/}} for better appreciation of the improved spectral reconstruction.

\section{Conclusions}

To address the magnitude over-attenuation in mid-to-high frequency regions, we propose two spectrally weighted $\mathcal{L}_\mathrm{Sig}$ and $\mathcal{L}_\mathrm{Adp}$, that modulate the phase-aware loss contribution as a function of frequency and signal characteristics. 
Experimental results show that HyST-Net serves as a competitive lightweight backbone for the proposed objective evaluations in streaming SE. Moreover, both losses yield consistent improvements in high-frequency reconstruction metrics, with $\mathcal{L}_\mathrm{Adp}$ further enhancing mid-frequency reconstruction through signal-dependent weighting while maintaining overall enhancement quality. 
Overall, the proposed spectrally adaptive loss achieves a more balanced spectral reconstruction across the full frequency range, alleviating the inherent trade-off between noise suppression and over-attenuation.

{\small
\setlength{\itemsep}{0pt}
\setlength{\parskip}{0pt}
\bibliographystyle{IEEEtran}
\bibliography{refs}
}

\end{document}